\documentstyle[12pt]{article}
\begin{document}

\begin{titlepage}
\baselineskip 20pt

\title{Heavy quark limit in the model with confined light quarks
and infrared heavy quark propagators}

\vspace{0.5cm}

\author{ M. A. Ivanov\thanks{Permanent address: Laboratory of Theoretical
Physics, Joint Institute for Nuclear Research, Head Post Office
P.O.Box 79, 101000 Moscow, Russia}
\,  and T. Mizutani\\
\\
\\
Department of Physics \\
Virginia Polytechnic Institute and State University\\
Blacksburg, VA 24061}
\maketitle

\vspace{1cm}

\abstract
\baselineskip 20pt

We have studied the weak decay constants and
the Isgur-Wise form factor of the B and D mesons in the heavy quark limit,
by employing a
relativistic quark confinement model. It is an attempt to improve our previous
work within the same line of thinking, but by incorporating a couple of novel
aspects. First,
the infrared behavior of the heavy quark is considered
by modifying its conventional propagator in terms of a single parameter
$\nu$. Second, the mass difference
of the heavy meson and heavy quark: $E=m_H-M_Q$ has been included.
It is found
that the weak decay constants depend strongly on the mass difference
E with a relatively mild $\nu$ dependence.
As for the Isgur-Wise function it is controlled more sensitively by the
infrared
parameter $\nu$, leading
to its suppression at maximum meson recoil.

\end{titlepage}

\section{Introduction}
\baselineskip 20pt

The theoretical treatment of {\it heavy-quark hadrons}: those containing a
heavy ($b$ or $c$) quark ($M_Q>>\Lambda_{QCD}$), is considerably simplified
due to a new {\it heavy quark} (spin-flavor) symmetry \cite{IW}.
Under this symmetry a heavy-quark meson may be pictured as
a system of a cloud of light degrees of freedom (light quarks and gluons)
surrounding the almost static color source (a heavy quark) whose four-momentum
being quite  close to its {\it on-mass-shell} value. In practice, it allows
for expressing all the $B\to D(D^*)l\nu$ decay form factors in terms of
a single universal form factor: the Isgur-Wise function (IWF), which is
determined primarily by the dynamics of the light degrees of freedom.

The systematical study of this symmetry is provided by the heavy-quark
effective theory (HQET) developed by Georgi {\it et al.} \cite{GEOR}.
In HQET
the heavy degrees of freedom are integrated out, while their effects as
virtual particles are taken into account by introducing non-renormalizable
higher-dimension operators proportional to the inverse powers of the
heavy quark mass \cite{FALK}. However, IWF cannot
be directly obtained within this scheme since it is essentially determined by
 the physics in the non-perturbative domain.
Thus, to work this quantity out
various quark models \cite{ISGW},\cite{N},\cite{IKM},\cite{HOLDOM},\cite{KARA},
\cite{MIT},\cite{BS},
QCD sum rules \cite{RAD},\cite{NEUB},\cite{BALL},\cite{PAVER},
and lattice calculations \cite{BERNARD}, \cite{UKQCD} have been employed so
far. The present work is devoted to the extension of our previous calculation
of IWF within a relativistic confined quark model \cite{IKM}. To constrain the
model parameters we will also calculate the weak decay constants of the $B$ and
$D$ mesons.
In order to set a basic stage for our study, we shall first briefly review
some of the model results to date in the following.

In the nonreltivistic confined quark model \cite{ISGW} IWF is
expressed in terms of the overlap integral of the initial and final meson
wave functions of the Gaussian form which are determined by fit to meson
spectra (note that heavy quark mass is assumed to be infinity here).
 With $w\equiv v\cdot v'$ where
$v(v')$ is the four-velocity of the initial (final) heavy-quark meson, it reads

\begin{equation}
\xi(w)=\exp\biggl\{-\rho^2(w-1)\biggr\}.
\end{equation}
The slope parameter obtained from the meson spectroscopy is
$\rho^2 =-\xi '(1) = 0.33$.
Anticipating that the result would be insufficient for large recoil, the
authors of \cite{ISGW} have introduced a universal reduction factor (from a
fit to the pion electromagnetic form factor) which has resulted in an increase
in the slope parameter: $\rho^2 \approx 0.64$. We note here a recent claim
\cite{WAM} that a due
consideration of the relativistic recoil correction and the Wigner rotation
of the spin of the light quark should increase the slope parameter to greater
than unity.

To improve certain aspects of nonrelativistic
approachs a simple relativistic-oscillator model with the light-front variables
was adopted to find IWF \cite{N},

\begin{equation}
\xi(w)=\frac{2}{w+1}\exp\{-(2\rho^2-1)\frac{w-1}{w+1}\}.
\end{equation}

\noindent Here, the slope parameter is related to the transverse momentum
distribution
of the light quark inside the heavy meson. By comparing with the data its value
has been extracted as $\rho \sim 1.14 \pm 0.23$.
\medskip

We next discuss other quark model approaches with or without due consideration
to confinement.
First, a simple model based on quark loop graphs
with an {\it ad hoc} form factor at each {\bf heavy-meson heavy-quark
light-antiquark}
vertex was proposed in Ref. \cite{HOLDOM}. It was
found that the physical quantities (including IWF) are quite sensitive
to the mass difference; $E \equiv m_H - M_Q$, of the heavy meson and heavy
quark
 when it exceeds $m_q$:
the light-quark (constituent) mass of the model. Then the mass function (self
energy),
decay form factor (IWF), etc. of the heavy meson
made up of a heavy-quark
light-antiquark loop develop  an imaginary part which becomes of the same
order of
magnitude as the real part. This is of course due to the absence of the quark
confinement in the model. Eventually the imaginary part has been discarded
completely by arguing that its neglect would mimic the quark confinement. To
date, the justification of such a procedure is not known.

A modified MIT bag model was applied to the calculation of IWF for the
ground state to ground state semi-leptonic decays involving the $b \to c$
transition \cite{MIT}.
The evaluation of the overlap integral yielded a result, which is claimed to
be well
approximated by the following form \cite{NEUB}

\begin{equation}
\xi(w)=\left[\frac{2}{w+1}\right]^{2+\frac{0.6}{w}}
\end{equation}.

The heavy-quark meson wave function was obtained by solving the bound state
Bethe-Salpeter equation with a modified one-gluon-exchange
ladder approximation \cite{BS}. The decay constant $f_B$ and IWF were then
calculated from the wave function. The slope parameter at zero recoil was
found to be rather large: $\rho^2=1.8-2.0$, whereas
the weak decay constant turned out very small, $f_B\approx 50$ MeV. Presumably,
the one-gluon-exchange may not be very realistic for generation of the heavy
meson bound state (this interaction alone does not confine the quarks).

Our previous approach for IWF \cite{IKM} was based upon the quark confinement
model (QCM) which incorporate the confinement of light quarks by devising
a quark propagator that has no singularities, thus forbidding the production of
a free quark \cite{EI}. This model then allows one to perform covariant
calculations
of Feynman diagrams with dressed light quark propagators free of simple
pole.  The calculation has given, in the heavy quark limit:
$M_Q=m_{H}\to\infty$, the IWF to decrease
slower than that obtained by many other
approaches except the bounds \cite{RAF} to be discussed below,

\begin{equation}
\xi(w)=\frac{1}{1+R}\left[\frac{2}{1+w}R+
\frac{1}{\sqrt{w^2-1}}\ln[w+\sqrt{w^2-1}]\right],
\end{equation}

\noindent
with the parameter fixed by the light quark confinement function (see the
next section) to be $R=0.65$. The corresponding
slope parameter
was extracted to be equal to $\rho^2=0.43$, quite small as compared with the
valued obtained from a fit to the data by the relativistic oscillater model as
discussed above. It should be mentioned that the IWF was obtained also with
finite heavy quark masses, and the result turned out to be very close to the
one in the heavy quark limit.

\medskip
It should be useful at this point to refer to a rather peculiar model
\cite{KARA} which, in part, has
motivated our present investigation.
It is based upon  the observation of the possible infrared
behavior of the heavy quark propagator. The physics here is not quite dynamical
but
mostly geometrical, i.e.
in terms infraparticle (heavy quark) propagation only without taking into
account
the light degrees of freedom. The obtained IWF
has the form

\begin{equation}
\xi(w)=1-\exp\biggl\{-\frac{1.02}{\sqrt{w^2-1}}\biggr\},
\end{equation}

\noindent
with the slope at zero recoil equal to zero. The model does not appear
particularly
realistic, but its consideration of the infrared propagator is quite
novel which we have adopted in our present work.

Different from various quark model approaches discused so far, the
QCD (spectral) sum rule \cite{SVZ} appears to be the only alternative so far in
effectively dealing the long distance physics of
the heavy mesons, aside from the direct numerical simulation by lattice
QCD.  This approach is based upon the hypothesis of quark-hadron duality in
the calculation of physical correlators and is combined with HQET to simplify
the physics.
After several
simplifying assumptions the first sum rule calculation
\cite{RAD} found

\begin{equation}
\xi(w)\approx \exp\{-0.37\sqrt{w^2-1}\}
\end{equation}
with the {\it infinite} \ slope parameter.

Later, this approach was improved \cite{NEUB} \cite{BALL} \cite{PAVER},
by including the {\it next to leading order} renormalization group
improvement, finite heavy-quark mass effects, etc.
It was found that these effects are quite large for
the weak decay constant and makes the D-meson as much as about $40\%$,
\cite{NEUB} causing a large deviation from the asymptotic scaling law
$f_D\sqrt{m_D}=f_B\sqrt{m_B}$. Also a rather strong dependence of this quantity
on the mass
difference $E=m_H-M_Q$ was observed .
As for IWF the E dependence was fairly weak or cancel out completely.
The main uncertainty brought in the calculation of this quantity comes from
the QCD perturbative estimate of the
higher-resonance
contributions to the spectral function in the dispersion integral
representation of the hadronic correlator. It turned out that the different
choice of the spectral function and of its integration domain produces
the values of the IWF form factor
at maximum recoil varing in the region $0.40\le\xi(w_{\rm max})\le 0.75$.
It was also shown that the infinite slope parameter of \cite{RAD} has resulted
from a na\" \i ve choice of the integration domain of the spectral function.
There appears, however, no single best choice for this domain.
The "best" result compared with the experimental data \cite{ARGUS}
may be parametrized as in eq.(3), giving $\rho^2=1.3$.

The lattice QCD has been exploited by two groups
\cite{BERNARD} and \cite{UKQCD} in the study of heavy meson physics.
The IWF was obtained by calculating the elastic D amplitude (or form factor)
$<D'|\bar c\gamma_{\mu} c|D>$,
in the quenched approximation. Fitting the result
to the relativistic-oscillator parametrization Eq.(2)
gives

\[
\rho^2=\left\{
\begin{array}{ll}
1.41\pm 0.19\pm 0.41 & \mbox{\cite{BERNARD}}\\
                     & \\
1.2^{+7}_{-3}        & \mbox{\cite{UKQCD}}
\end{array}
\right.
\]

Lastly, we should mention that the bounds on the IWF were studied without
relying on particular models, but only with
the
consideration of analyticity
and dispersion relation for the heavy meson elastic form factor assuming the
heavy quark symmetry and the hardon-quark duality
\cite{RAF}. It was claimed that the bounds is very strongly constrained.
These bounds prohibit the IWF from falling by less than 12$\%$
and more than 21$\%$ from zero to maximum recoil. The lower bound conflicts
with many model predictions. It has been argued by \cite{WISE},
\cite{GRIN}, \cite{ISGUR}, and \cite{KOERN} that the careful inclusion of heavy
quark-antiquark bound states, lying below the heavy meson pair production
threshold, may gives no such small lower bound. The authors of \cite{RAF}
recently included the
effect of the $\Upsilon$ states below the $B \bar B$ threshold \cite{RAF2},
claiming
that the rigorous upper bound for the IWF slope at zero recoil is now
$\rho^2 \approx 6.0$, although, on phenomenological grounds, it
may be diminished to
$\rho^2 \le 1.7$.

The main goal of the present paper is to improve our previous work \cite{IKM}
as we
believe that our QCM approach possesses a sound dynamical background regarding
its modelling of
confined light quarks, successful in many static and non-static
properties of the light hadrons \cite{EI}. Thus the present investigation will
focus on how one should treat the behavior of the heavy quark within
the context of this QCM.

\section{Quark propagators}
\baselineskip 20pt

The quark confinement model (QCM) \cite{EI} is specified by the interaction
Lagrangian inferred from the partial hadronization of the DCQ action.
Then the transition between heavy meson $H$ and heavy $Q$
and light $q$ quarks is described by

\begin{equation}
{\cal L}_H(x)=g_H H(x)\bar Q(x)\Gamma_H q(x)
\end{equation}
with the coupling constants $g_H$ defined by what is usually called the
$compositeness \, \, \, condition$ which means that the renormalization
constant of the meson field is equal to zero:

\begin{equation}
Z_H=1-g^2_H\Pi_H^\prime(m^2_H)=0.
\end{equation}
Here $\Pi_H^\prime$ is the derivative of the meson mass operator.
The compositeness condition also provides the right normalization
of the charge form factor $F_H(0)=1$. This could be readily seen from
the Ward identity

\begin{equation}
g^2_H\Pi_H^\prime(p^2)=g^2_H{1\over 2p^2}p^\mu
{\partial \Pi_H(p^2)\over\partial p^\mu}=g^2_H{1\over 2p^2}p^\mu
T^\mu_{HH}(p,p)=F_H(0)=1.
\end{equation}
with the three-point function $T^\mu(p,p^\prime)$ defined as

\begin{equation}
T^\mu_{HH}(p,p^\prime)={3\over 4\pi^2}\int{d^4k\over 4\pi^2i}
{\rm tr}\biggl\{\Gamma S_Q(\not\! k-\not\! p^\prime)
\gamma^\mu S_Q(\not\! k-\not\! p)\Gamma S_q(\not\! k)
\biggr\}.
\end{equation}

The matrix elements describing the leptonic decays $H(H^*)\to e\nu$ ($H =
B,D$),
and semileptonic ones $B\to D(D^*)e\nu$ are written down as

\begin{equation}
M^\mu \{ H(H^*) \to e\nu \}=g_H T^\mu_H(p),
\end{equation}
where

\begin{equation}
T^\mu_H(p)={3\over 4\pi^2}\int{d^4k\over 4\pi^2i}
\cdot{\rm tr}\biggl\{O^\mu S_Q(\not\! k-\not\! p)
\Gamma S_q(\not\! k)
\biggr\},
\end{equation}
and

\begin{equation}
M^\mu_{B\to D(D^*)e\nu}=g_Bg_{D(D^*)} T^\mu_{HH^\prime}(p,p^\prime),
\end{equation}
where

\begin{equation}
T^\mu_{HH^\prime}(p,p^\prime)={3\over 4\pi^2}\int{d^4k\over 4\pi^2i}
{\rm tr}\biggl\{\gamma^5 S_{Q^\prime}(\not\! k-\not\! p^\prime)
O^\mu S_Q(\not\! k-\not\! p)\Gamma S_q(\not\! k)
\biggr\}.
\end{equation}
\\

Now we discuss the forms of quark propagators.

\vspace{0.5cm}
\noindent
{\bf (1) Confined light-quark propagator.}
\vspace{0.3cm}

Here we briefly explain the salient feature of QCM concerning the light-quark
propagator. The principal assumption \cite{EI} is that the propagator is
an  entire analytical function in the complex-"$\not\! p$ \ " plane with the
{\it constituent mass}
$m_q\equiv z\Lambda$ being smeared by complex measure $d\sigma_z$ such that

\begin{equation}
\int{d\sigma_z\over\Lambda z-\not\! p}={1\over\Lambda}
\biggl[a\left(-{p^2\over\Lambda^2}\right)+{\not\! p\over\Lambda}
b\left(-{p^2\over\Lambda^2}\right)\biggr]
\end{equation}
with the confinement functions $a(u)$ and $b(u)$ defined by

\begin{equation}
a(u)=\int d\sigma_z {z\over z^2+u^2} \hspace{1cm}
b(u)=\int d\sigma_z {1\over z^2+u^2}.
\end{equation}
The scale parameter $\Lambda$ characterizes the size of the confinement region.
It has turned out \cite{EI} that the low-energy physics depends
only on those quantities which involve the integrals of $a(u)$ and $b(u)$
together with $u^\alpha$ $\;$  $(\alpha\ge0)$ but not on the detailed shape
of these functions. We have devized a simple choice of the confinement
functions
\cite{EI}:

\begin{equation}
a(u)=a_0\exp(-u^2-a_1u) \hspace{1cm} b(u)=b_0\exp(-u^2+b_1u).
\end{equation}
The parameters $a_i$, $b_i$, and $\Lambda$ have been determined from the best
model description of hadronic properties at low energies and the following
values were found:

\begin{eqnarray}
& &a_0=b_0=2, \hspace{1cm} a_1=1, \hspace{1cm} b_1=0.4,\nonumber\\
& &\\
& &{\rm and} \hspace{1cm} \Lambda=460 {\rm MeV},\nonumber
\end{eqnarray}
which describe various physical observables quite well \cite{EI}.

\vspace{0.5cm}

\noindent
{\bf (2) Infrared heavy-quark propagator.}

\vspace{0.3cm}

Since a heavy quark (with mass $M_Q$) in a heavy meson is under the influence
of soft gluons (which sets the scale $\Lambda_{QCD} << M_Q$), it
may be regarded as nearly on its mass-shell where
the infrared regime should take place for its
propagation. The infrared behavior for one-fermion Green's function
(propagator) has been investigated in various papers (see, for instance,
\cite{KARA}, \cite{BARB},and
the references therein).
The result is well-known only for abelian gauge theories:

\begin{equation}
S(p)\sim (m-\not\! p-i\epsilon)^{-1-\nu},
\end{equation}
where $\nu=(\alpha_S/4\pi)(3-\lambda)$ with $\lambda$ being the gauge
parameter.
It seems quite resonable to see how sensitive the physical observables may be
to this modification even though the integrals in Eqs.(10), (12), (14) have no
infrared divergences. To make our present study simple, we choose the
heavy quark propagator as in Eq.(19),

\begin{equation}
S_Q(k+p)=(M_Q-\not\! k-\not\! p)^{-1-\nu}
\end{equation}
assuming that $\nu$ is a free parameter. Using the conventional
notation for the heavy meson four-momentum $p=(M_Q+E)v$, with $v$ being the
four-velocity
of the heavy meson (or of the heavy quark since there is a velocity
selection rule: \cite{GEOR}, one readily finds that this propagator satisfies
the form necessary for the correct heavy quark limit:

\begin{equation}
S_Q(k+p)=\left[1\over M_Q-\not\! k-\not\! p \right]^{1+\nu}
={1\over 2}{1+\not\! v\over \left(-E-kv\right)^{1+\nu}}+O({1\over M_Q}).
\end{equation}
The physical observables are defined by two parameters: the heavy meson
heavy quark mass difference
$E = m_H - M_Q$ which typifies the binding of the light and heavy quarks,
and $\nu$ characterizing
the infrared behavior of the heavy quark as disscussed above. It should be
mentioned here that in our previous work \cite{IKM} we have not considered the
effect of the non-vanishing $E$. In the heavy quark limit both the meson
mass $m_H$ and heavy quark mass $M_Q$ are sent to infinity, so it does not
appear to make sense to keep this value finite. However, it characterizes
the binding of the light and heavy quarks, thus
an very important parameter particularly in constraining the values
of the weak decay constant, as we shall see later.
We also
note that this quantity plays a significant role in other approaches.

\section{Weak decay constants and IW form factor}
\baselineskip 20pt

The calculation of the two- and three-point functions (see Eqs.(10),(12), and
(14)) is considerably simplified by taking only the leading order in the
$1/M_Q$ expansion (the heavy quark limit) since
the calculations of the traces and the integrations over internal momentum $k$
factorize in this limit due to the following identities:

\begin{eqnarray}
& &{\rm tr}\left[O^\mu(1+\not\! v)\Gamma (z+\not\! k)\right]=
{\rm tr}\left[O^\mu(1+\not\! v)\Gamma (z+\not\! v kv)\right]=
{\rm tr}\left[O^\mu(1+\not\! v)\Gamma\right] \left(z-kv\right)\\
& &\nonumber\\
& &{\rm tr}\left[\Gamma(1+\not\! v)\gamma^\mu(1+\not\! v)\Gamma
(z+\not\! k)\right]=
2v^\mu{\rm tr}\left[\Gamma\Gamma\right] \left(z-kv\right)\\
& &\nonumber\\
& &{\rm tr}\left[\Gamma(1+\not\! v^\prime)O^\mu(1+\not\! v)\gamma^5
(z+\not\! k)\right]\\
&=&{\rm tr}\left[\Gamma(1+\not\! v^\prime)O^\mu(1+\not\! v)\gamma ^5
\left(z+{kv+kv^\prime\over 2(w+1)}(\not\! v+\not\! v^\prime)-
{kv-kv^\prime\over 2(w-1)}(\not\! v-\not\! v^\prime)\right)\right]
\nonumber\\
&=&{\rm tr}\left[\Gamma(1+\not\! v^\prime)O^\mu(1+\not\! v)\gamma ^5\right]
\left(z-{kv+kv^\prime\over (w+1)}\right)\nonumber
\end{eqnarray}
where $w=v\cdot v^\prime$.
Here we have taken into account that $\{\Gamma,\not\! v\}=0$ for both
$\Gamma=\gamma^5$ and $\gamma^\nu$, which follows from the transversality of
the vector field.

Using these identities in the Eqs.(10), (12) and (14) gives

\begin{eqnarray}
T^\mu_{HH}(p,p)&=&2p^\mu \left({\Lambda\over m_H}\right)
\left({1\over 2\Lambda^\nu}\right)^2
{3\over 4\pi^2}J_3^{(\nu)}({E\over\Lambda},1)\cdot L_0,\\
\nonumber\\
T^\mu_{H}(p)&=&\Lambda\left({\Lambda\over m_H}\right){1\over 2\Lambda^\nu}
{3\over 4\pi^2}J_2^{(\nu)}({E\over\Lambda})\cdot L_1^\mu,\\
\nonumber\\
T^\mu_{HH^\prime}(p,p^\prime)&=& {\Lambda\over(2\Lambda^\nu)^2}
{3\over 4\pi^2}J_3^{(\nu)}({E\over\Lambda},w)\cdot L_2^\mu.
\end{eqnarray}
Here we use notation:

\[L_0=\left\{
\begin{array}{ll}
I                 & \mbox{($\Gamma=\gamma^5$)}\\
                  &                         \\
g^{\alpha\beta}   & \mbox{($\Gamma=\gamma^\alpha$)}
\end{array}
\right. \]

\[L_1^\mu=\left\{
\begin{array}{ll}
p^\mu                   & \mbox{($\Gamma=\gamma^5$)}\\
                        &                          \\
M\epsilon^\mu         & \mbox{($\Gamma=\gamma^\alpha$)}
\end{array}
\right. \]

\[L_2^\mu=\left\{
\begin{array}{ll}
(v+v^\prime)^\mu & \mbox{($\Gamma=\gamma^5$)}\\
                 &                          \\
i\varepsilon^{\mu\epsilon v^\prime v}+v^{\prime\mu}(v\epsilon)-
\epsilon^\mu(w+1) & \mbox{($\Gamma=\gamma^\alpha$)}
\end{array}
\right. \]
where $\epsilon$ is the polarization vector.
The structural integrals $J_2^{(\nu)}$ and $J_2^{(\nu)}$ are defined as

\begin{eqnarray}
J_2^{(\nu)}(E)&=&\int {d^4k\over\pi^2i}
\int d\sigma_z {z-kv\over(z^2-k^2)(-E-kv)^{1+\nu}}\\
& &\nonumber\\
J_3^{(\nu)}(E,w)&=&\int{d^4k\over\pi^2i}
\int d\sigma_z
{
z-(kv+kv^\prime)/(w+1)
\over
(z^2-k^2)(-E-kv)^{1+\nu}(-E-kv^\prime)^{1+\nu}
}
\end{eqnarray}
and they are calculated in the Appendix.

The $HQ\bar q$ coupling constant is determined from Eq.(8) and may be written
down as

\begin{equation}
g(m_H)=2\Lambda^\nu{4\pi\over\sqrt{3}}\sqrt{m_H\over\Lambda}
\sqrt{1\over J_3^{(\nu)}(E/\Lambda,1)}.
\end{equation}

Finally, the matrix elements of the leptonic and semileptonic decays
are written as

\begin{eqnarray}
M^\mu_{H(H^*)\to e\nu}&=&f_H\cdot L^\mu_1 \\
& &\nonumber\\
M^\mu_{B\to D(D^*)e\nu}&=&\sqrt{m_Bm_D}\xi(w)\cdot L_2^\mu
\end{eqnarray}
with the weak decay constant $f_H$ and the IWF given by

\begin{eqnarray}
f_H\sqrt{m_H}&=&\Lambda^{3/2}{\sqrt{3}\over 2\pi}
{J_2^{(\nu)}(E/\Lambda)\over\sqrt{J_3^{(\nu)}(E/\Lambda,1)}},\\
\nonumber\\
\xi(w)&=&{J_3^{(\nu)}(E/\Lambda,w)\over J_3^{(\nu)}(E/\Lambda,1)}.
\end{eqnarray}

Now we discuss the dependence of physical observables on parameters
$E$ and $\nu$. Note that hereafter $E$ is given in units of $\Lambda$
(= 460 MeV),
At this point, it may be
useful to stress that unlike in Ref. \cite{HOLDOM} there is no imaginary part
in our matrix elements for any value of E  since there is no $Q\bar q$
production threshold due to confinement. The value of E is not arbitrary,
however.
The first restriction comes from the requirement that the integral
$J_3^\nu(E,1)$ in Eq.(30) should be a positive definite quantity. Since
this integral for $\nu>0$ is defined in terms of the derivatives of the
confinement
functions, it develops zeros above certain values of E. The behavior of
$J_3^\nu(E,1)$
for  $\nu=0, 0.5,1$ is shown in Fig.1. Another restriction on
E comes from the value of the heavy meson weak decay constants.
{}From eq.(33) it is clear that our result in the heavy quark limit obeys
the scaling relation $f_B\sqrt{m_B}=f_D\sqrt{m_D}$.
The dependence of the weak decay constant values on E is
shown in Fig.2, and it is clear that this quantity is very
sensitive to the change in E: strongly increasing function of E. This is in
line with what was observed in QCD sum rule approaches, see for example,
\cite{NEUB}. In the present result the E=0 case is unrealistically small,
suggesting that even in the heavy quark limit E must be kept finite as it
should be.

The values of the weak decay
constants obtained in different models are in variance but
eventually $f_D$ never exceeds $\sim 250$ MeV \cite{NEUB}, \cite{DOMI},
\cite{LATT}, \cite{LAT}, \cite{LAT1} when the pion  decay constant is
normalized to $f_\pi=$132 MeV.
We will use the above observations as a guide to constrain E.
The numerical results for the upper bound obtained in this way for various
choice of the infrared parameter $\nu$ are shown in Fig.3. It is interesting
to note the lower bound for E obtained in Ref. \cite{GURA}
E$\ge 0.5$, which might eventually be
implemented to get the restriction on the infrared parameter $\nu$ which turns
out as $\nu\le 0.75$. Note, however, that there is an antithesis to the way
this
lower bound has been derived \cite{BIGI}.

The IWF is plotted in Fig.4 for various values of E and
$\nu$. One can see that the E-dependence
of this function is much weaker than that seen in the weak decay
constants. This agrees qualitatively with the result obtained by using
QCD sum rules
\cite{NEUB}, \cite{BALL}, \cite{PAVER}. Note, in particular, that in
\cite{NEUB} the
IWF is {\it completely} independent of E, which is rather curiuous intuitively.
In the present result
the E-dependence decreases for increasing $\nu$: for $\nu=0$ and at maximum
recoil the difference between
the values of the IWF for E=0 and E=1.0 is no greater than $12\%$. The general
tendency is that the IWF is suppressed for larger values of E.

On the other hand, the dependence on
$\nu$ is much stronger as seen in Fig.5.
For examlpe, at maximum recoil  the IWF is suppressed by 13$\%$ for
$\nu=0.5$, and by 23$\%$ for $\nu=1$ in comparison with $\nu=0$ when E=0.
A similar tendency is observed for other values of E.

The E-dependence of the slope parameter at zero recoil:
$\rho^2=-\xi^\prime(1)$, which is an important quantity characterizing the IWF,
is shown in Fig.6. One can see that it is contained within the following
region:

\begin{eqnarray}
& &0.42\le\rho^2\le 0.64 \hspace{1cm} (\nu=0),\\
& &\nonumber\\
& &0.64\le\rho^2\le 0.82 \hspace{1cm} (\nu=0.5),\\
& &\nonumber\\
& &0.62\le\rho^2\le 0.64 \hspace{1cm} (\nu=1).
\end{eqnarray}

The result is consistent with the {\it generous} Bjorken lower bound:
$\rho^2\ge 0.25$
\cite{BJOR}: but somewhat less than the values obtained in other approaches
(see, Table 2). It should be noted, however, that a recent fit to the
CLEO $B \to D^* l\nu$ data with various analytic forms adopted for the IWF
give $\rho^2 \approx 1.1 \pm 0.5 \pm 0.3$, consistent with our present result
\cite{BESS} with $\nu \approx 0.5$. Also note
that this quantity does not necessarily determine the
global behavior of the IWF, as seen in Fig.8. Also, its accurate determination
from experimental data is not trivial, in good part, due to the small count
near zero recoil. The effort is being made from the CLEO data for
the $B \to D^*l \bar \nu$ and we await a quick release of the analysis.

Plot of the IWF for the set of parameters $\{\nu,E\}$ providing
the reasonable values for the weak decay constants (see, Table 1.) is
presented in Fig.7. One can see that, for example, the curve with the
parameters
$\nu=0.75$, E=0.5 (or E=230 MeV) is consistent with the experimental result
within $19\%$ off the best fit taken from Ref. \cite{STONE}.
In Fig.8 the IWF taken from the other approaches are shown.

\section{The effects of finite masses in the weak decays}
\baselineskip 20pt

To study the correction to the heavy quark limit, we will take into
account the effect of finite heavy quark masses in the weak decays of B and
D mesons.
This is of special interest since the QCD sum rule \cite{NEUB}, \cite{DOMI}
and the lattice calculations \cite{LATT}, \cite{LAT} give an indication
of a strong breakdown of the scaling law:

$$f_D=170\pm 30 \, {\rm MeV} \hspace{1 cm} f_B=190\pm 50 \, {\rm MeV}
\hspace{1cm} \cite{NEUB}$$

$$f_D=150 \to 195 \,{\rm MeV} \hspace{1 cm} f_B=150 \to 185 \, {\rm MeV}
\hspace{1cm} \cite{DOMI}$$

$$f_D=200\pm 25 \, {\rm MeV} \hspace{1 cm} f_B=230\pm 10 \, {\rm MeV}
\hspace{1cm} \cite{LATT}$$

$$f_D=185^{+4+42}_{-3-7}  \, {\rm MeV} \hspace{1 cm}
  f_B=160^{+6+53}_{-6-19} \, {\rm MeV}
\hspace{1cm} \cite{LAT}$$

$$f_D=208(9)\pm 35\pm 12  \, {\rm MeV} \hspace{1 cm}
  f_B=187(10)\pm 34\pm 15 \, {\rm MeV}
\hspace{1cm} \cite{LAT1}$$

In \cite{DOMI} the spread in the value of the decay constants
originates from varying the values of the heavy quark masses.
Also the results are found without adopting any HQET, and the sum rule result
with the method of Hilbert moments not listed here
has given somewaht different values. Note that in \cite{NEUB} and \cite{LATT}
the complete break down of the scaling law even brings the value of $f_D$ lower
than that of $f_B$.

To simplified the calculations, we consider only the case with $\nu=0$ in
view of its very mild influence on the decay constants discussed earlier.
Then we have

\begin{equation}
f_H\sqrt{m_H}=\Lambda^{3/2}\frac{\sqrt{3}}{2\pi}
\frac
{J_2^{(0)}(M_Q/\Lambda,E/\Lambda)}
{\sqrt{J_3^{(0)}(M_Q/\Lambda,E/\Lambda)}}
\end{equation}
with the structural integrals defined as

\begin{eqnarray}
J_2^{(0)}(M,E)&=&4\int\limits_0^\infty\frac{dtt}{(1+t/M)^3}
\biggl[t(1+\frac{t}{2M})-E(1+\frac{E}{2M})\biggr]
\biggl\{a(z)+\frac{1}{2}\frac{t}{1+t/M} b(z)\biggr\}\nonumber\\
& &\\
J_3^{(0)}(M,E)&=&4\int\limits_0^\infty\frac{dtt}{(1+t/M)^3}
\biggl\{a(z)+\frac{t}{2}
\biggl[3+\frac{t}{M}-\frac{(1+E/M)^2}{1+t/M}\biggr]
b(z)\biggr\},\nonumber
\end{eqnarray}
where

$$z=\frac{t}{1+t/M}\biggl\{t-2E(1+\frac{E}{2M})\biggr\}.$$
It is easy to show that eq.(33), whose structural integrals
being defined in the Appendix, is reproduced in  the heavy quark limit
$M_Q\to\infty$.

We have adopted the values of the heavy mesons: $m_D = 1. 87$ GeV, $m_B=5.27$
GeV, and varied $E=m_H-M_Q$ which, in turn, determines the finite heavy quark
masses. Just like in the heavy quark limit as discussed earlier, $f_H$ has
turned out
quite small for small E, and the effect of the finite $M_Q$ is found
rather insignificant. However, for $E \ge 0.5$ \ $f_H$ increases significantly.
In order to compared with the result in the heavy quark limit as listed in
Table 1, we we present here the finite quark mass result for E=1:

\begin{equation}
f_D=175 \, {\rm MeV} \hspace{1cm} f_B=125 \, {\rm MeV},
\end{equation}
which is quite reasonable as compared with the lattice and QCD sum rule
results quated above except that the B-meson decay constant is somewhat on
the smaller side. When compared with our value in the heavy quark limit,
the reduction is about 10$\%$ for the B meson and as large as 25$\%$ for the
D meson. So obviously this effect is important and breaks the scaling law.
The percentage reduction here is in agreement with what was found in
ref.\cite{NEUB}
within the QCD sum rule when the extra reduction factor $(M_Q/m_H)^{1.5}$ in
this work is left out. We should note that it is this extra factor in this
reference that eventually reverses the magnitudes of $f_D$ and $f_B$ (note
that there the adopted value of E is somewhat smaller for D), thus consistent
with the lattice result in \cite{LATT}. In our present approach this reduction
factor is apparently absent: we wonder if it is a model independent factor we
have missed, or it is due to the specific way the parameters were defined in
performing the Laplace sum rule in the heavy quark limit in \cite{NEUB}. In
any event, we should stress that the present section is not intended as
predicting the values of the heavy meson decay constants: rather it is to
demonstrate the importance of the finite heavy quark mass effect to
correct the heavy quark limit as long as this specific physical quantity is
concerned, particularly for the D meson.

As for the finite quark mass effect on the IWF, we already studied it in our
previous work \cite{IKM} where $E\equiv 0$. The effect was found to be
rather small and we expect a similar conclusion to hold here, thus has not
been investigated.

\newpage
\section{Conclusion}
\baselineskip 20pt

As an extension of our previous work \cite{IKM}, we have presented a quark
model approach to the decay of heavy quark mesons where a confined light quark
and the near on-shell infrared behavior of the heavy quarkpropagation are
taken into account. The infrared behavior of the heavy quark is modelled in
terms of a single parameter $\nu$ which modifies the simple free Feynman
propagator. The study has been carried out in the heavy quark limit:
$M_Q \to \infty$. However, we have retained the finite value for
$E = m_H -M_Q$: the difference between the heavy meson and heavy quark masses.
We have found that the weak decay constant of the heavy quark mesons depends
strongly on E while its $\nu$-dependence is rather mild. This fact has then
been
used to constrain the value of E by imposing a conservative upper bound for
$f_H$ to be $\approx 250$ MeV for the D meson decay. In this way it is found
that for a reasonable value $E \sim 460$ MeV with $\nu$ =0 ( corresponding to
a free heavy quark propagation) the value $f_D$ is found to be consistent with
approaches like QCD sum rule \cite {NEUB}, \cite {DOMI} and a lattice
simulation \cite{LATT}, \cite{LAT}, \cite{LAT1}.

In the same context we have studied the universal semi-leptonic decay form
factor: the Isgur-Wise function. Unlike in the case of the decay constants, it
is found to be controlled more dominantly by the infrared parameter $\nu$
which gives a desirable suppression of this function for $\nu > 0$. As an
example, for a set of parameters: E=0 and $\nu=0.75$ giving
{\it reasonable}
values for the weak decay constants, the IWF turns out to be consistent
with the best empirical fit to the ARGUS data \cite{ARGUS} nearly within
the experimental
error bars: we may need a little extra suppresion.

Our treatment of the infrared behavior of the
heavy quark propagator in terms of a single parameter may seem too simple.
It is motivated by studies in the
abelian gauge field, thus is not rooted from real QCD. However, we believe
that it
certainly points to the direction of further investigation, as it appears
reasonable enough to believe that a free heavy quark propagator (with a simple
pole at the quark mass) may not be realistic but needs to be modified in such
a way to respect the heavy quark symmetry.

We have also studied the effect of the finite heavy quark mass, and found that
it gives a significant deviation from the scaling law for the weak decay
constants obtained in the heavy quark limit (and also in the non-relativistic
quark models). We thus confirmed the claims from certain lattice and QCD sum
rule approaches.

To conclude, we want to reiterate the following: while the importance of
retaining the finite value for E appears to be established in other approaches,
the possible significance of the infrared behavior of the heavy
quark propagator has not been paid much attention except in a simple model
\cite{KARA}.  While we ourselves are in persuit of a more realistic form of the
heavy quark infrared propagator, we
will certainly feel happy if the present work may invoke
certain interest in this direction.

\vspace{0.5in}

\begin{center}
{\bf ACKNOWLEDGMENTS}
\end{center}

We  would like to thank N.Isgur, A.Radyushkin, J. K\"{o}rner, T. Mannel and
M. Neubert for useful discussions.
This work was supported in part by the United States Department of Energy
under Grant No. DE-FG05-84-ER40413.

\newpage

\begin{center}
{\bf APPENDIX}
\end{center}

\noindent
{\bf Two-point function.}

\begin{eqnarray}
J_2^{(\nu)}(E)&=&\int {d^4k\over\pi^2i}
\int d\sigma_z {z-kv\over(z^2-k^2)(-E-kv)^{1+\nu}}
\nonumber\\
& &\nonumber\\
&=&{\Gamma(2+\nu)\over\Gamma(1+\nu)}
\int\limits_0^1 d\alpha (1-\alpha)^\nu
\int{d^4k\over\pi^2i}
\int d\sigma_z
{z-kv\over \biggl[\alpha(z^2-k^2)-(1-\alpha)(E+kv)\biggr]^{2+\nu}}
\nonumber\\
& &\nonumber\\
&=&{\Gamma(2+\nu)\over\Gamma(1+\nu)}
\int\limits_0^\infty dt t^\nu
\int\limits_0^\infty duu\int d\sigma_z
{z+t/2\over \biggl[z^2+u-tE+t^2/4\biggr]^{2+\nu}} \nonumber
\end{eqnarray}

\noindent
(1) $0\le \nu<1$

\begin{eqnarray}
& &J_2^{(\nu)}(E)=2^\nu{\sin{\pi \nu}\over \pi \nu}
\int\limits_0^\infty du u^{(1-\nu)/2}
\int\limits_0^1 dx{x^{(\nu-1)/2}\over (1-x)^\nu}
\left\{a(u-2E\sqrt{xu})+\sqrt{xu}b(u-2E\sqrt{xu})\right\}.\nonumber
\end{eqnarray}

\noindent
(2) $\nu=1$

\begin{eqnarray}
& &J_2^{(\nu)}(E)=2\int\limits_0^\infty du
\left\{a(u-2E\sqrt{u})+\sqrt{u}b(u-2E\sqrt{u})\right\}.\nonumber
\end{eqnarray}

\noindent
{\bf Three-point function.}

\begin{eqnarray}
J_3^{(\nu)}(E,w)&=&\int{d^4k\over\pi^2i}
\int d\sigma_z
{z-(kv+kv^\prime)/(w+1)\over(z^2-k^2)(-E-kv)^{1+\nu}(-E-kv^\prime)^{1+\nu}}
\nonumber\\
& &\nonumber\\
&=&{\Gamma(2+2\nu)\over\Gamma^2(1+\nu)}
\int\limits_0^1 d\beta \beta^\nu(1-\beta)^\nu
\int{d^4k\over\pi^2i}\int d\sigma_z
{
z-(kv+kv^\prime)/(w+1)
\over
(z^2-k^2)\biggl[-E-k(\beta v+(1-\beta) v^\prime)\biggr]^{2+2\nu}
}
\nonumber\\
& &\nonumber\\
&=&2^{1+2\nu}{\Gamma(1+2\nu)\over\Gamma^2(1+\nu)}
\int\limits_0^1 d\tau {\tau^\nu(1-\tau)^\nu\over W^{1+\nu}}
\int\limits_0^\infty duu^\nu\int d\sigma_z
{z+\sqrt{u/W}\over \biggl[z^2+u-2E\sqrt{u/W}\biggr]^{1+2\nu}} \nonumber
\end{eqnarray}
with W given by

$$
W=1+2\tau(1-\tau)(w-1).
$$

\noindent
(1) $\nu=0$

\begin{eqnarray}
& &J_3^{(\nu)}(E)=2\int\limits_0^1{d\tau\over W}\int\limits_0^\infty du
\left\{a\left(u-2E\sqrt{u/W}\right)+
\sqrt{u/W}b\left(u-2E\sqrt{u/W}\right)\right\}.\nonumber
\end{eqnarray}

\noindent
(2) $\nu=0.5$

\begin{eqnarray}
& &J_3^{(\nu)}(E)={16\over\pi}\int\limits_0^1 d\tau{\sqrt{\tau(1-\tau)}
\over W^{3/2}}
\int\limits_0^\infty du\sqrt{u}
\left\{a_1\left(u-2E\sqrt{u/W}\right)+
\sqrt{u/W}b_1\left(u-2E\sqrt{u/W}\right)\right\}.
\nonumber
\end{eqnarray}

(3) $\nu=1$

\begin{eqnarray}
& &J_3^{(\nu)}(E)=8\int\limits_0^1 d\tau{\tau(1-\tau)\over W^{2}}
\int\limits_0^\infty duu
\left\{a_2\left(u-2E\sqrt{u/W}\right)+
\sqrt{u/W}b_2\left(u-2E\sqrt{u/W}\right)\right\},
\nonumber
\end{eqnarray}
where $f_i(u)=(-)^if^{(i)}(u)$, $(f=a,b)$.

\noindent
(4) $0<\nu<0.5$

\begin{eqnarray}
J_3^{(\nu)}(E)&=&2^{1+2\nu}{\sin{2\pi \nu}\over 2\pi \nu}
{\Gamma(1+2\nu)\over\Gamma^2(1+\nu)}
\int\limits_0^1 d\tau {\tau^\nu(1-\tau)^\nu\over W^{1+\nu}}
\int\limits_0^1 dx    {x^\nu\over (1-x)^{2\nu}}
\int\limits_0^\infty duu^{1-\nu}\nonumber\\
& &\nonumber\\
& &\left\{a_1\left(u-2E\sqrt{ux/W}\right)+
\sqrt{ux/W}b_1\left(u-2E\sqrt{ux/W}\right)\right\}.
\nonumber
\end{eqnarray}

\noindent
(4) $0.5<\nu<1$

\begin{eqnarray}
J_3^{(\nu)}(E)&=&{2^{2\nu}\over \nu}{\sin{\pi(2\nu-1)}\over \pi (2\nu-1)}
{\Gamma(1+2\nu)\over\Gamma^2(1+\nu)}
\int\limits_0^1 d\tau {\tau^\nu(1-\tau)^\nu\over W^{1+\nu}}
\int\limits_0^1 dx    {x^\nu\over (1-x)^{2\nu-1}}
\int\limits_0^\infty duu^{2-\nu}\nonumber\\
& &\nonumber\\
& &\left\{a_2\left(u-2E\sqrt{ux/W}\right)+
\sqrt{ux/W}b_2\left(u-2E\sqrt{ux/W}\right)\right\}.
\nonumber
\end{eqnarray}

\newpage
\listoffigures

\noindent
Fig.1. The dependence of the structural integral $J_3^\nu(E,1)$ (in arbitrary
units) on E for
various choices of the infrared parameter $\nu$.

\noindent
Fig.2. The dependence  of the weak decay constants on E for
various choices of the infrared parameter $\nu$.

\noindent
Fig.3. The correlation of E and $\nu$ which provides
the weak decay constants: $f_D\sim 250$ MeV,
and $f_B\sim 150$ MeV.

\noindent
Fig.4. The IW function for various choices of E with the fixed value of
infrared
parameter $\nu$.

\noindent
Fig.5. The IW function for various choices of the infrared parameter
$\nu$  with fixed E.

\noindent
Fig.6. Slope parameter (ar zero recoil) as a function of E for
various choices of the infrared parameter $\nu$.

\noindent
Fig.7. Plot of the IW function for the set of parameters $\{\nu,E\}$ providing
the reasonable values for the weak coupling constants (see, Table 1.).
The data points are taken from Ref.\cite{ARGUS}.
The dash curve represents the best fit of $e^{-\rho^2(w-1)}$ with
$\rho^2=1.18\pm 0.50$ taken from Ref. \cite{STONE}. The dot curves
represent the upper and lower bounds of the fit that related to
the experimental uncertainties.

\noindent
Fig.8. Plot of the IW functions taken from the other approaches:
VQM (valence quark model) \cite{ISGW}; QCD (QCD sum rules) \cite{RAD};
ROM (relativistic quark model) \cite{N};
INFRA (infrared heavy quark propagator) \cite{KARA};
The data points are taken from Ref.\cite{ARGUS}.
FIT (the best fit of experimental data) \cite{STONE}.

\newpage
\listoftables

\noindent
Table 1. The magnitudes of weak decay constants and slope parameter
of IW function for some set of parameters $\{\nu,E\}$.

\noindent
Table 2. The slope parameter of IW function in other approaches including our
previous work \cite{IKM}.

\newpage
\begin{center}
{\large Table 1.\\}
\end{center}
\def\arraystretch{2.0}
\begin{center}
\begin{tabular}{|c|c|c|c|}
\hline\hline
 & $\nu=0.00$ & $\nu=0.50$ & $\nu=0.75$ \\
 &  E=460 {\rm MeV} & E=276 {\rm MeV} & E=230  {\rm MeV} \\
\hline
$f_D$ ({\rm MeV}) & 231  & 216 & 226 \\
\hline
$f_B$ ({\rm MeV}) & 137  & 140 & 135 \\
\hline
$\rho^2$          & 0.61 & 0.83 & 0.92\\
\hline\hline
\end{tabular}
\end{center}

\begin{center}
{\large Table 2.\\}
\end{center}
\def\arraystretch{2.0}
\begin{center}
\begin{tabular}{|l|l|}
\hline\hline
Bjorken \cite{BJOR} & $>\frac{1}{4}$ \\
\hline
Isgur \cite{ISGW}    & 0.63(0.33) \\
\hline
Rosner \cite{ROSN}   & $1.44\pm 0.41$ \\
\hline
Mannel \cite{MANN} & $1.77\pm0.74$ \\
\hline
Neubert \cite{NEUB} & $1.28\pm 0.25$ \\
\hline
Bernard \cite{BERNARD} & $1.41\pm 0.19\pm 0.41$ \\
\hline
UKQCD Collab. \cite{UKQCD} & $1.2^{+7}_{-8}$ \\
\hline
Radyushkin \cite{RAD} & $\infty$ \\
\hline
Karanikas \cite{KARA} & 0 \\
\hline
Sadzikowski \cite{MIT} & 1.24 \\
\hline
Kugo \cite{BS} & 1.8-2.0 \\
\hline
Ivanov {\it et al.} \cite{IKM} & 0.43 \\
\hline\hline
\end{tabular}
\end{center}

\end{document}